\documentstyle[aps,twocolumn,epsbox]{revtex}
\begin{document}
\draft
\newcommand{\oo}{\o}
\newcommand{\simj}{\stackrel{>}{_\sim}}
\newcommand{\simk}{\stackrel{<}{_\sim}}
\title
{
Superconductivity  of the   One-Dimensional  $d$-$p$ Model with $p$-$p$ transfer
}
\author{
Kazuhiro {\sc Sano}   and  Yoshiaki {\sc \=Ono}\raisebox{0.5ex}{\dag}
}
\address{
 Department of Physics Engineering,  Mie University, Tsu, Mie 514.       \\ \raisebox{0.5ex}{\dag}Department of Physics, Nagoya University, Nagoya 464-01
}
\date{\today}
\maketitle
\begin{abstract}
  Using the numerical diagonalization method, we investigate the one-dimensional $d$-$p$ model, simulating a Cu-O linear chain with strong  Coulomb repulsions.  
Paying attention to   the effect of the transfer energy $t_{pp}$ between the nearest neighbor oxygen-sites,  we calculate  the critical exponent of correlation functions $K_{\rho}$ based on the Luttinger liquid relations and the ground state energy $E_0(\phi)$ as a function of an external flux $\phi$.
We find that the transfer $t_{pp}$ increases the charge susceptibility and  the exponent $K_{\rho}$ in cooperation with  the repulsion  $U_{d}$ at Cu-site. We also show  that  anomalous flux quantization occurs for $K_{\rho}>1$.
The superconducting region is presented on a phase diagram of $U_{d}$ vs. $t_{pp}$ plane.
\end{abstract}
 Low-dimensional  strongly correlated electron systems    attract much interest due to the possible relevance to high-$T_c$ superconductivity.
A considerable number of  theoretical studies have been performed on the strongly correlated fermion models  either in one-band or multi-band version to understand the mechanism of the high temperature  superconductivity.
 However, there  is  no consensus on what the actual mechanism of the  superconductivity is,  except for the fact that  the strong repulsive interaction on Cu-site plays an important role in the mechanism. \cite{Anderson,Varma1} 
To achieve a solid understanding of this problem, some rigorous studies, even of the simplest models, are highly desirable at this stage. For such purpose, numerical diagonalization studies of  finite size  systems may be successful.\cite{Hirsch,Ogata2,Dopf2,Ohta} 
In particular, one-dimensional $d$-$p$ model, simulating a Cu-O linear chain with strong  Coulomb repulsions, provides a good target to investigate the mechanism of the high-$T_c$ superconductivity.\cite{SanoPhysica1,SanoPRB,SanoPhysica2,Barford,Sudbs,Stechel,Sandvik,Zotos}
The model contains hopping $t_{pd}$ between Cu($d$)-site and oxygen($p$)-site and repulsive interaction at the $d$- and $p$-site ($U_d$ and $U_p$, respectively). In addition, it can contain  the nearest-neighbor $d$-$p$ interaction $U_{pd}$ and/or  hopping  $t_{pp}$ between the  nearest-neighbor $p$-sites.

In the previous work,\cite{SanoPhysica1,SanoPRB,SanoPhysica2} the present authors have studied the  one-dimensional $d$-$p$ model with large on-site Coulomb repulsion  $U_d$  at Cu-sites and inter-site repulsion $U_{pd}$,  by using the numerical diagonalization method.  
   With the help of  the Luttinger liquid relations,  the superconducting(SC) correlation is found to be  dominant compared with the CDW and SDW correlations  in the proximity of  the phase separation. 
 The SC phase appears for $1.0\simk U_{pd}\simk 1.6$  almost independent of  the filling with  $1.3 \simk n\simk 1.7$.  Near the half-filling $n=1$ and the full-filling $n=2$, however, the SC correlation is suppressed. \cite{Barford}
 Recently, Sudb\oo $\ $ et al. showed that the  $d$-$p$ chain with large $U_{pd}$ exhibits flux quantization with charge $2e$ and slow algebraic decay of the singlet SC correlation function on oxygen sites.\cite{Sudbs}   
 Other studies of the  $d$-$p$ chain  also showed that   $U_{pd}$  enhances the charge fluctuation as well as the SC fluctuations and when it exceeds a certain critical value the system becomes unstable towards  phase separation.\cite{Stechel,Sandvik}  
 All of these studies  claimed that the parameter $U_{pd}$ is central in inducing a SC state at $T=0$. 

However, if one adds  the hopping term $t_{pp}$ to the  model, the situation will be changed completely.  It  enhances the charge fluctuation and increases   the exponent $K_{\rho}$  as the parameter $U_{pd}$ does. Therefor, the repulsive interaction  $U_{pd}$ is not always necessary for the SC state.
   In the present paper, to clarify the effect of $t_{pp}$, we study  the  one-dimensional $d$-$p$ model with  the hopping $t_{pp}$  by the numerical diagonalization.
We calculate the critical exponent of correlation functions $K_{\rho}$  and the ground state energy $E_0(\phi)$ as a function of an external flux $\phi$.
Using the results for the critical exponent, we show a SC region on the  $U_{d}-t_{pp}$ plane.

We consider the following model Hamiltonian for the  Cu-O chain:

\begin{eqnarray} 
  H&=&t_{pd}\sum_{<ij>,\sigma} (p_{i\sigma}^{\dagger} d_{j\sigma}+h.c.)
 +t_{pp}\sum_{<ij>,\sigma} (p_{i\sigma}^{\dagger} p_{j\sigma}+h.c.)  \nonumber \\
  &+&\epsilon_{d}\sum_{j,\sigma} d_{j\sigma}^{\dagger}d_{j\sigma}     +\epsilon_{p}\sum_{i,\sigma} p_{i\sigma}^{\dagger}p_{i\sigma}
    + U_d\sum_{j}n_{dj\uparrow}n_{dj\downarrow}
%   + U_p\sum_{i}n_{pi\uparrow}n_{pi\downarrow}
%    + U_{pd}\sum_{<ij>,\sigma \sigma' }n_{pi\sigma}n_{dj\sigma'}     , 
\end{eqnarray} 
where $d^{\dagger}_{j\sigma}$ and $p^{\dagger}_{i\sigma}$ stand for creation operators of a hole with spin $\sigma$ in  the Cu$(d)$-orbital  at site $j$ and of a hole with spin $\sigma$ in the O$(p)$-orbital at site $i$ respectively, and 
   $n_{dj\sigma}=d_{j\sigma}^{\dagger}d_{j\sigma}$.
 Here,  $t_{pd}$ stands for the transfer energy  between the nearest neighbor $d$- and $p$-sites, which will be set to be unity ($t_{pd}$=1) hereafter in the present study.  
 The  atomic energy  levels of $d$- and $p$-orbitals are given by  $\epsilon_{d}$ and $\epsilon_{p}$, respectively. The charge-transfer energy $\Delta$ is defined as $\Delta=\epsilon_{p}-\epsilon_{d}$. 

  To  achieve systematic calculation, we use the periodic boundary  condition for $N_h=4m+2$ and antiperiodic boundary condition for $N_h=4m$, where $N_h$ is the total hole number and m is  an  integer. This choice of the boundary condition removes accidental degeneracy so that the ground state might always be a singlet with zero momentum. 
 The filling $n$ is defined  by  $n=N_{h}/N_{u}$, where $N_u$ is the total number of unit cells ( each unit cell contains a $d$- and a $p$-orbitals) and the Fermi wave number  $k_F$ is given as $k_F=\frac{\pi }{2}n$.
 
 We numerically diagonalize the Hamiltonian  up to 12 sites (6unit  cells)  using the standard Lanczos algorithm. 
   
   The chemical potential $\mu (N_{h},N_{u})$ is defined by
\begin{equation}
   \mu (N_{h},N_{u})=\frac{E_{0}(N_{h}+1,N_u)-E_{0}(N_{h}-1,N_u)}2\quad, 
\end{equation}
where $E_{0}(N_h,N_u)$ is the total ground state energy of a system with $N_u$ unit cells and $N_h$ holes.
     When the charge gap vanishes in the thermodynamic limit, the uniform  charge susceptibility $\chi_c$ is obtained from 
\begin{equation}
 \chi_c(N_{h},N_{u})=\frac{2/N_u}{\mu(N_{h}+1,N_{u})-\mu(N_{h}-1,N_{u})}  
\end{equation}

In the Luttinger liquid theory,  some relations  have been  established as universal relations in the one-dimensional  models.\cite{Haldane,Haldane2,Voit}  
 In  the model which is isotropic in spin space,  the critical exponents of various types of correlation functions are determined by a single parameter $K_\rho$, which is the exponent of the power-law decay of  correlation functions.
For the one-dimensional  $d$-$p$  model in the weak coupling regime,  Matsunami and  Kimura \cite{Matsunami}  calculated the critical exponents of some correlation functions by using the renormalization group analysis ($g$-ology), and shown that the system belongs to the Tomonaga-Luttinger (TL) liquid.
In the present authors' recent  work,\cite{SanoPRB} we have  shown  the numerical results of the critical exponent $K_\rho$ and the phase diagram on the $U_d$-$U_{pd}$ plane  including the  weak coupling region and the strong coupling region.
In the weak coupling region, it agrees with the results of the $g$-ology.  An  analysis of spin-gap  indicates  that the strong coupling region also belongs  to the TL regime and  not to the Luther-Emery (LE) regime. Here, the LE regime is characterized by a gap of the spin excitation spectrum, while in the TL regime, the excitation  is gapless. \cite{Voit,Solyom}

 It is  predicted that SC correlation  is dominant for $K_{\rho}>1$ (the correlation function decays   as $\sim r^{-(1+\frac{1}{K_{\rho}})}$ in the TL regime and as $\sim r^{-\frac{1}{K_{\rho}}}$ in the LE regime), whereas Charge Density Wave (CDW)  or Spin Density Wave ( SDW) correlations is dominant  for $K_{\rho}<1$ (correlation functions decay  as $\sim r^{-(1+K_{\rho})}$ in the TL regime and as $\sim r^{-K_{\rho}}$ in the LE regime).
In the TL regime, the singlet SC and triplet SC correlation functions have the same critical exponent apart from a logarithmic correction.
On the other hand, in the LE regime,  the  correlation functions of SDW and triplet SC decrease exponentially.   
The parameter $K_{\rho}$ is related to the charge susceptibility $\chi_c$ and the charge velocity $v_c$ by the relations, \cite{Ogata2,Haldane,Voit,Schulz}
\begin{equation}
          K\sb{\rho}=\frac{\pi}{2}v\sb{c}\chi\sb{c} 
\end{equation}
with $v_c=\frac{N_u}{2\pi}(E_{1}-E_{0})$ ,where $E_{1}-E_{0}$ is the lowest charge excitation energy.
  We can also determine the $K_{\rho}$ by the Drude weight $D$
\begin{equation}
      K\sb{\rho}=\frac{1}{2}(\pi \chi_c D)^{1/2} 
\end{equation}
with $D=\frac{\pi}{N_u} \frac{\partial^2 E(\phi)}{\partial \phi^2}$, where $E(\phi)$ is the total energy of the ground state as a function of flux 
$\phi$.\cite{Voit}
Using these two independent equations of the $K_{\rho}$, we will  check the consistency of the Luttinger liquid relations.   

\begin{figure}
%\figureheight{13cm}
\begin{center}
\psbox[height=4.8cm]{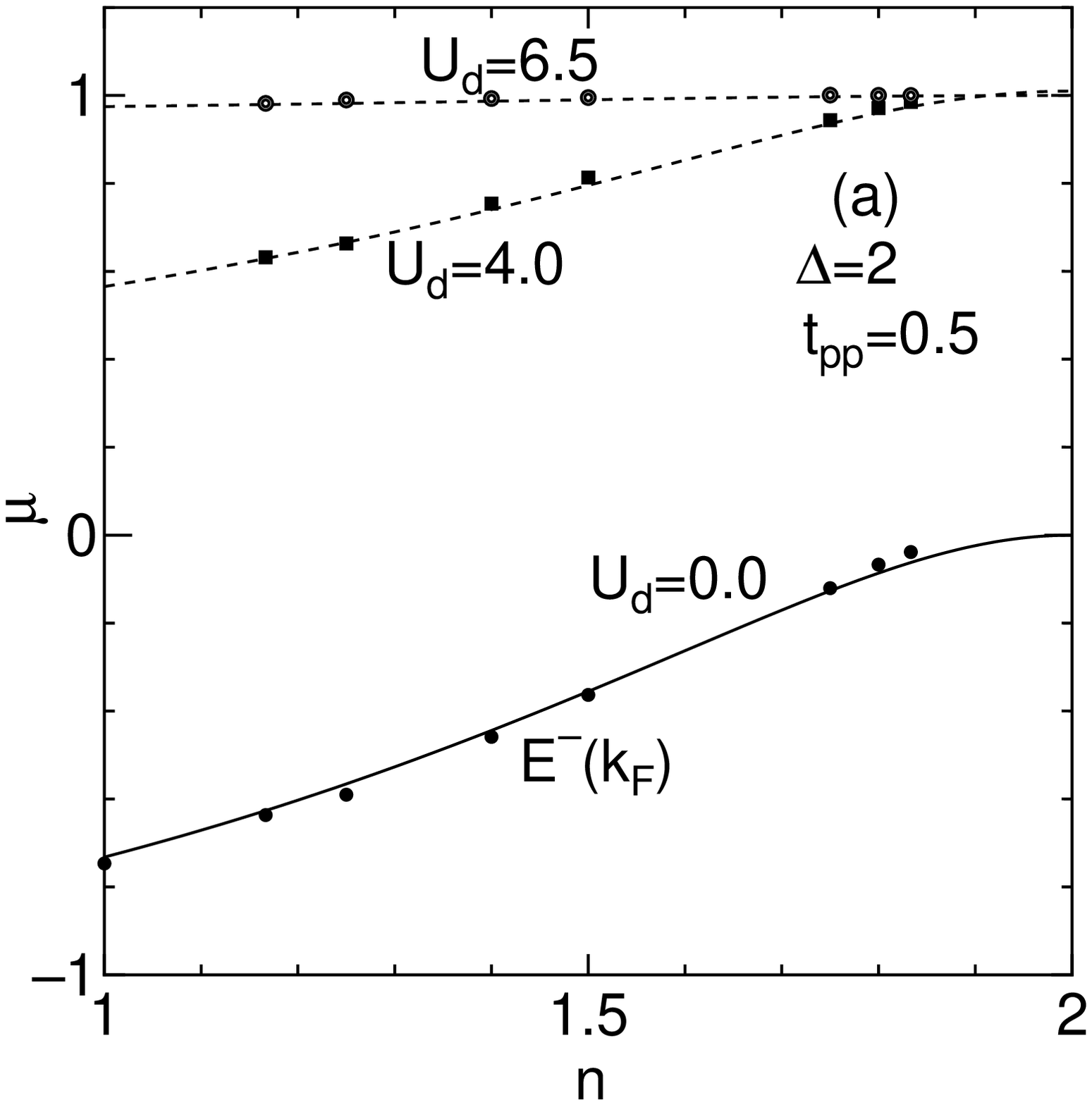}
\psbox[height=4.8cm]{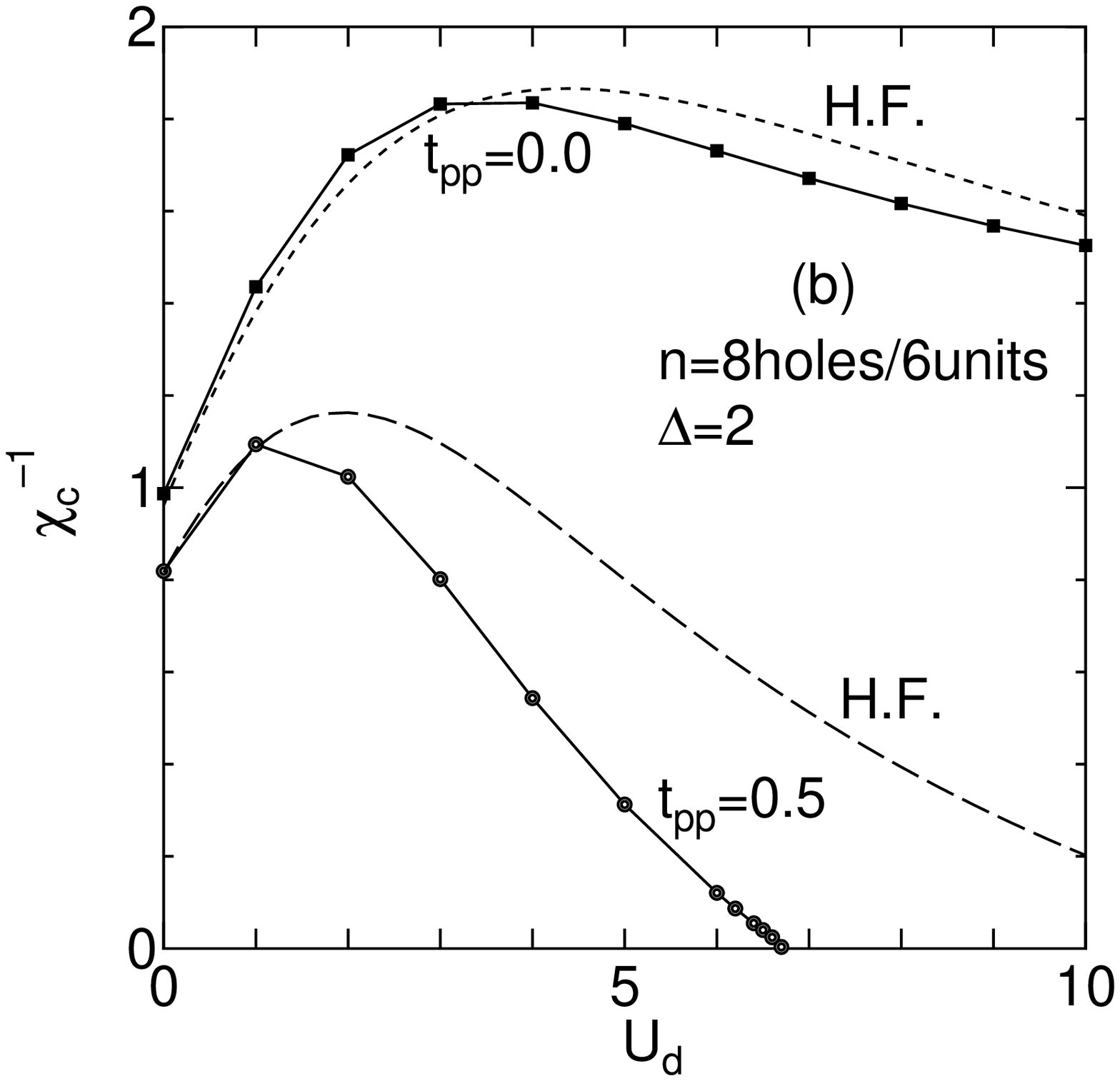}
\end{center}
\caption{
(a) The chemical potential $\mu$ as a function of the filling $n$ for various values of $U_d$. 
 $\mu$  is calculated for 
$N_u$= 8, 10 and  12. 
The  solid line represents  $E^-(k_F)$. The broken lines are guide for eye.
(b) The inverse charge susceptibility $\chi^{-1}$ as a function of  $U_d$   for $t_{pp}=0.0$ and 0.5 with the results in the H.F. approximation. 
}
\label{fig:1}
\end{figure}

For a non-interacting system ($U_d=0$), the Hamiltonian (1) is easily diagonalized and a dispersion relation is   given as 

\begin{eqnarray} 
 E^{\pm}(k)&=&\frac{1}{2} \Bigl\{ \epsilon_{d}+\epsilon_{p}+2t_{pp}\cos{k}  \nonumber \\
&\pm& \sqrt{(\Delta+2t_{pp}\cos{k})^2+16(t_{pd}\cos{(k/2)})^2} \Bigr\}    \label{band}
\end{eqnarray} 
   where $k$ is a wave vector and $ E^+(k)( E^-(k) )$ represents a upper (lower) band energy. 
Compared with the case with $t_{pp}=0$, the width of the lower band $ E^{-}(k)$ becomes narrower for $t_{pp}>0$. Note that  it decreases with $\Delta$ and becomes  perfectly flat at $\Delta=2t_{pp}-t_{pd}^2/t_{pp}$. When  one decreases $\Delta$ further, the band bends with a peak at $k=0$. 
To investigate the band structure of interacting systems, we consider the chemical potential $\mu$, which is corresponding to Fermi energy $E_F$ at  $T=0$.  
Figure \ref{fig:1}(a) shows $\mu$ as a function of the filling $n$ for several values of $U_d$ with $\Delta=2$ and $t_{pp}=0.5$. 
We calculate the value of $\mu$ by using eq.(2)  at the hole densities $n=5/4, 7/4,7/5,9/5,,7/6,9/6$ and $11/6$.    Data points with  $U_d=0$  seem to sit closely along on the dispersion curve of $E^-(k_F)$. It suggests that the finite-size effect is sufficiently small.

When  $U_d$ is large, $\mu$ is large and weekly depends  on $n$.
Especially, the slope of $\mu$   becomes almost flat at $U_d=6.5$. 
In this case, the charge susceptibility $\chi_c$ almost diverges, because of the relation,  $\chi_c^{-1}=\frac{d\mu}{dn}$. 
The result suggests that the width of effective hole band becomes very narrow due to a strong correlation effect of large repulsion $U_d$. 
In order to see the behavior of the charge susceptibility  more qualitatively, in Fig.~\ref{fig:1}(b), we show the inverse of the charge susceptibility $\chi^{-1}$ as a function of $U_d$ in a system with 6 unit cells for $\Delta=2$. 
In the case of $t_{pp}=0$, the $U_d$-dependence of $\chi^{-1}$ is week and  $\chi^{-1}$ remains as a finite value. On the other hand,  it is close to zero at $U_d \simeq 6.75$ for $t_{pp}=0.5$. 
It shows that the transfer $t_{pp}$ increases the charge susceptibility in cooperation with  the repulsion  $U_{d}$.
If we use the Hartree-Fock(H.F.) approximation,  the  atomic  levels of $d$-orbitals  $\epsilon_{d}$ are pushed up as $\tilde{\epsilon_{d}}=\epsilon_{d}+\frac{1}{2}U_d<n_d>$, where $<n_d>$  is the mean value of hole density $n_d$ at a $d$-site and
 determined by solving   the self-consistent equation of $<n_d>$.
Then, the band structure is approximately described as a renormalized band $\tilde{E}(k)$ whose $\Delta$ is replaced by $\tilde{\Delta}=\Delta-\frac{1}{2}U_d<n_d>$. 
The band width of  
$\tilde{E}^-(k)$ decreases as $\tilde{\Delta}$ decreases and  becomes zero at sufficiently large $U_d$. The decreases of the band width leads to the enhancement of $\chi_c$ as shown in Fig.1(b).

Next, we  show  the critical exponent of correlation functions $K_{\rho}$ as a function of $U_d$ in Fig.2(a).
We calculate  $K_{\rho}$ in two ways by using eq.(4) and eq.(5) independently, one is obtained by combining the charge velocity $v_c$ and the charge susceptibility $\chi_c$ and the other is expressed by the Drude weight.
The result shows that the values of $K_{\rho}$ obtained by in the two different ways  are well consistent with each other except for $U_d\simk-1$.
We also show $K_{\rho}$ obtained by the $g$-ology together in Fig.2(a).
 In the weak coupling regime, the results of numerical diagonalization are in good agreement with  those of the $g$-ology.
These results indicate that   the Luttinger liquid theory can be applied to the one-dimensional $d$-$p$ model successfully  and  finite-size effect of $K_{\rho}$  is not so large.

To judge which of the two  estimations of $K_{\rho}$ is better from a point of view of  finite-size effect, we watch the values of $K_{\rho}$ at $U_d=0$. The one from eq.(4) is equal to 0.972 and the other from eq.(5) is 1.002. Taking account of $K_{\rho}$ being always equal to unity in noninteracting case, we infer that
 the estimation  from  eq.(5) is better than the other. 
Furthermore, the comparison of those values of  $K_{\rho}$ with the  values obtained in the $g$-ology in the weak coupling regime also indicates that the estimation of  $K_{\rho}$ from eq.(5)  gives a better result. 
Therefore, we will use the value of $K_{\rho}$ expressed by eq.(5) in the following.

Fig.2(a) shows that $K_{\rho}$  decreases as $U_d$ increases when $U_d\simk 3$.  However,   $K_{\rho}$ increases with $U_d$ for $U_d\simj 3$  and  diverges at $U_d\sim 6.8$  where  the charge susceptibility diverges. 
The divergence indicates that the width  of the effective band  $\tilde{E}^-(k)$ is close to zero. 
When $K_{\rho}$ is larger than unity, the SC correlation is expected to be most dominant compared with CDW and SDW correlations. 
The region where $K_{\rho}> 1$ appears at $6.5\simk U_d \simk 6.8$. It seems to be fairly small, but  surely exists.
To confirm  the ground state to be in the SC state, we show the ground state energy $E_0(\phi)$ as a function of an external flux $\phi$ in Fig.2(b). The  anomalous flux quantization occurs at $U_d=6.7$, where $K_{\rho}$ is about 2.0. When $U_d=6.0$, $K_{\rho}$ is less than  unity and  the anomalous flux quantization is not found.
These results  support that the SC phase  really appears at $K_{\rho}> 1$.

Generally speaking, the SC phase is induced by an attractive interaction between quasi-particles. To examine the effective attraction, we introduce the repulsive interactions $U_{dd}$ ($U_{pp}$) between the nearest-neighbor $d$-$d$ ($p$-$p$) sites and the repulsion $U_p$.
If those repulsions act between  a relevant pair of holes or cancel the  effective attraction, the SC state is expected to be suppressed and $K_{\rho}$ is reduced. 
We find that $U_{dd}$ does not affect  $K_{\rho}$, but  $U_{p}$ ($\simj0.2$), as well as $U_{pp}$,  decreases it. 
It suggests that  the relevant pair of holes sit not on the $d$-sites  but on the $p$-sites and   the strength of the attraction  is of  order of $0.2$.\cite{Matsukawa}
\begin{figure}
%\figureheight{13cm}
\begin{center}
\psbox[height=4.8cm]{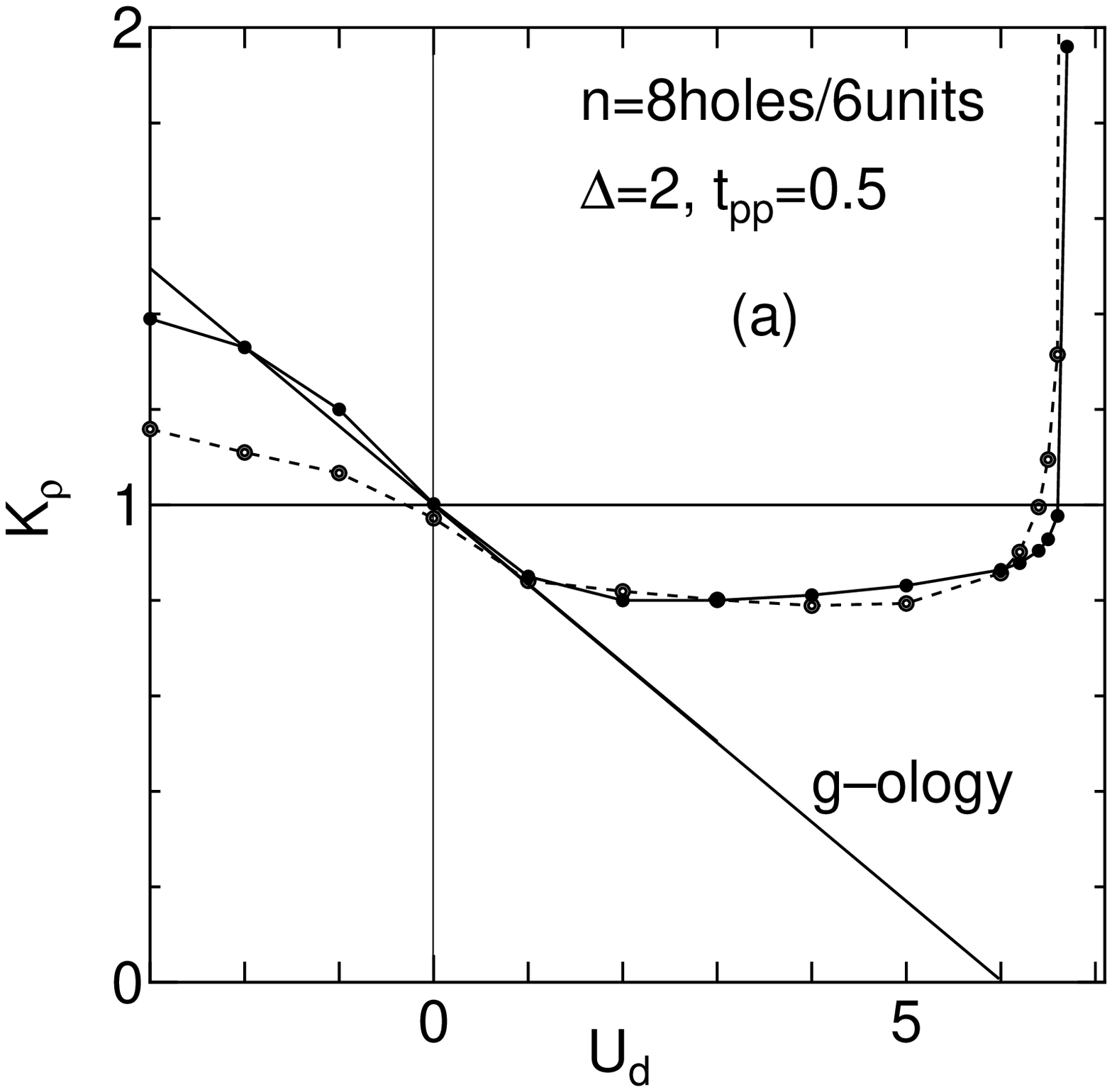}
\psbox[height=4.8cm]{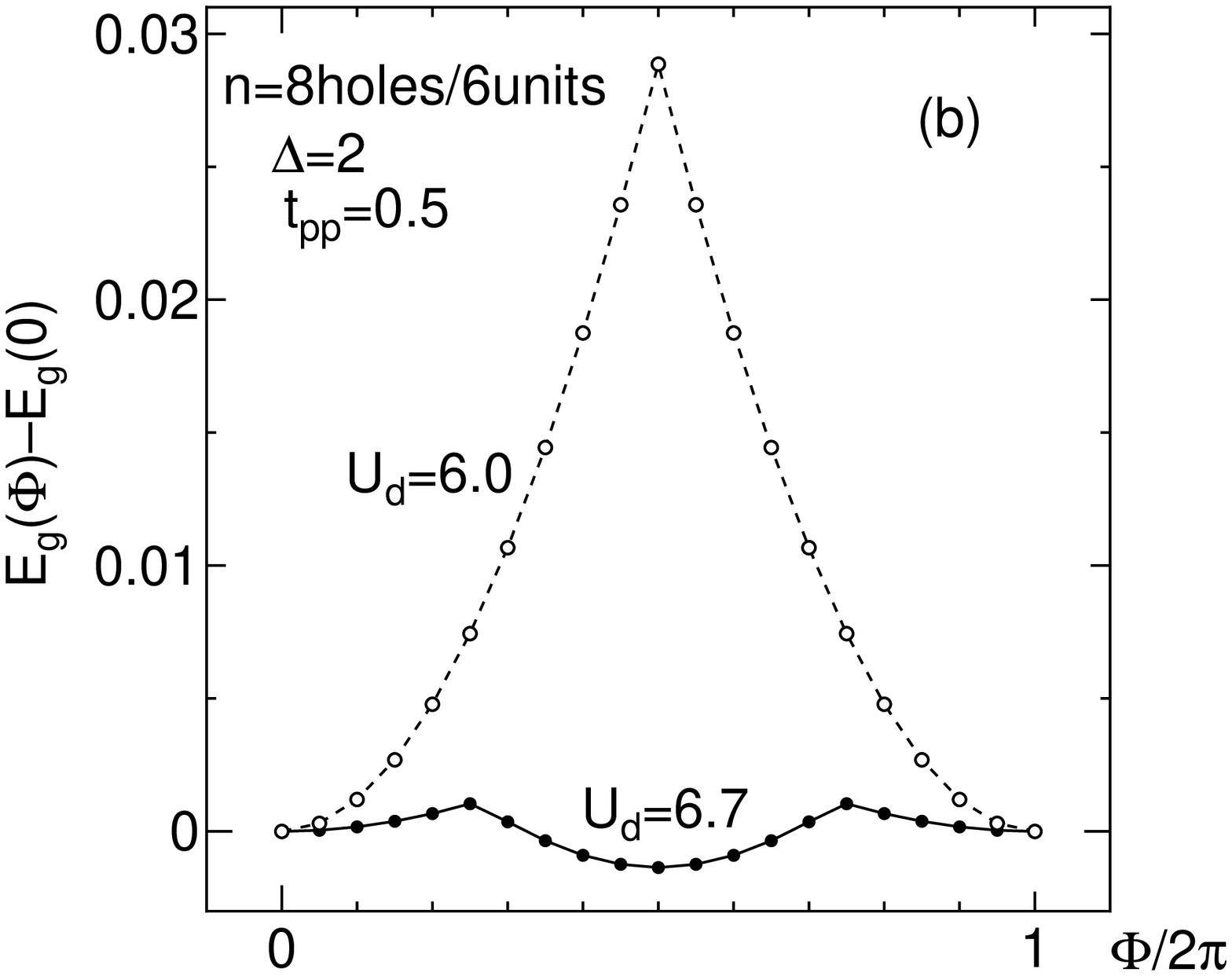}
\end{center}
\caption{
(a) $K_\rho$ as a function of $U_{d}$.  The solid circles with the solid line represent the estimation  from eq.(5) and the open circles with the broken line represent the estimation from eq.(4). (b) The energy difference  $E_0(\phi)-E_0(0)$ as a function of an external flux$\phi$. 
}
\label{fig:2}
\end{figure}
\begin{figure}
%\figureheight{6.5cm}
\begin{center}
\psbox[height=5cm]{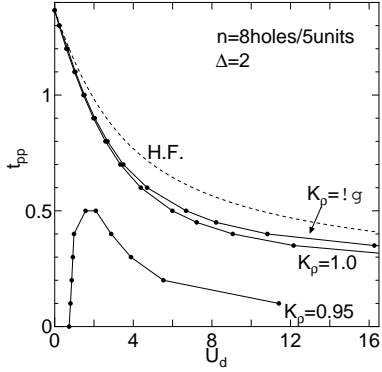}
\end{center}
\caption{
 Contour map for the $K_\rho$ on the $U_d$-$t_{pp}$ plane  at  $\Delta=2$  for  the 5-unit  8-hole system ($n$=8/5).
 The broken line represents $t^c_{pp}$ obtained in the H.F. approximation. 
}
\label{fig:3}
\end{figure}
 Finally, we investigate the critical exponent $K_{\rho}$ in detail and obtain the phase diagram in the $U_d-t_{pp}$ parameter  plane. 
   Figure  3 shows a  contour map of $K_{\rho}$ for $n=8/5$ system. 
  The  region sandwiched between two lines with  $K_{\rho}=1$ and   $K_{\rho}=\infty$ is corresponding to the SC region. The region is very small for small $U_d$, however, it increases with $U_d$.

In the H.F. approximation,  the renormalized band $\tilde{E}^-(k)$ becomes flat at a critical value $t^c_{pp}$ which is given as 
\begin{eqnarray}
 t^c_{pp}=\frac{1}{4} \Bigl\{ \Delta&-&\frac{1}{2}U_d<n_d> \nonumber \\
  &+& \sqrt{(\Delta-\frac{1}{2}U_d<n_d>)^2+8t_{pd}^2} \Bigr\}. \nonumber
\end{eqnarray}
Here, the approximation breaks down for $t_{pp} \geq t^c_{pp}$.
 We expect that the point of $t_{pp}=t^c_{pp}$ corresponds to that of  $K_{\rho}=\infty$. 
In the noninteracting system ($U_d=0$),  $t^c_{pp}=\frac{1+\sqrt{3}}{2}$ for $\Delta=2$.
As $U_d$ increases,  $t^c_{pp}$ decreases as shown in Fig.3. It seems that the line of $t^c_{pp}$ is near  that of $K_{\rho}=\infty$.
  
In summary, we have numerically diagonalized the one-dimensional $d$-$p$  model with finite sizes.  Paying special attention to the role of $t_{pp}$, we have calculated chemical potential and critical exponent $K_{\rho}$. We have also showed the  anomalous flux quantization observed in the SC state.
Using  the Luttinger liquid relations, we have found that the SC correlation is  dominant compared with the CDW and SDW correlations  in the proximity of the parameter region where the renormalized band becomes flat.
  We have confirmed that  $t_{pp}$  enhances the charge  fluctuation and promotes the SC correlation as  $U_{pd}$ does.  

\section*{Acknowledgement}
  The authors  would like to thank  Y. Kuroda  for  critical reading  of our manuscript.
This work has been supported partially by the Grant-in-Aid for Scientific Research on Priority Area from the Ministry of Education, Science, Sports and Culture, and also by CREST(Core Research for Evolutional Science and Technology) of Japan Science and Technology Corporation(JST).
 The computation in the present work has been partly done using the facilities of the Supercomputer Center, Institute for Solid State Physics, University of Tokyo.

% \begin{figure}
% \caption{}
% \label{}
% \end{figure}


\begin{references}


%
\bibitem{Anderson}
 P.W. Anderson, Science {\bf 235}, 1196 (1987). 
%
\bibitem{Varma1}
 C.M. Varma, S. Schmitt-Rink and E. Abrahams, Solid State Commun. {\bf 62}, 681 (1987). 
%

\bibitem{Hirsch}
 J.E. Hirsch, E. Loh, Jr., D.J. Scalapino and S. Tang, Phys. Rev. B {\bf 39}, 243 (1989). 
%
%
\bibitem{Ogata2}
M. Ogata, M. U. Luchini, S. Sorella and F. F. Assaad, 
Phys. Rev. Lett. {\bf 66},  2388 (1991). 
%
\bibitem{Dopf2}
G. Dopf, A. Muramatsu and W. Hanke, Phys. Rev. Lett. {\bf 68},  353 (1992).
\bibitem{Ohta}
 Y. Ohta, T. Shimozato, R. Eder and S. Maekawa,  Phys. Rev. Lett, {\bf 73},  324 (1994). 
%
\bibitem{SanoPhysica1}
K. Sano and Y. \={O}no, Physica {\bf C205}, 170 (1993).
%
\bibitem{SanoPRB}
K. Sano and Y. \={O}no, Phys. Rev. {\bf B51},  1175 (1995).
%
\bibitem{SanoPhysica2}
K. Sano and Y. \={O}no, Physica {\bf C242},  113 (1995).
%
\bibitem{Barford}
W. Barford and E. R. Gagliano Physica {\bf B194-196},  1455 (1994); see also, C. Vermeulen, W. Barford and E. R. Gagliano, Europhys. Lett. {\bf 28},  653 (1994).
%
\bibitem{Sudbs}
 A. Sudb\oo, C.M. Varma, T. Giamarchi, E.B. Stechel, and T. Scalettar,  Phys.  Rev.  Lett. {\bf 70}, 978 (1993). 
%
%
\bibitem{Stechel}
 E.B. Stechel, A. Sudb\oo, T. Giamarchi, and C.M. Varma,  Phys.  Rev.  {\bf B51},  553 (1995). 
%
\bibitem{Sandvik}
 A. W. Sandvik, A. Sudb\oo,  Phys.  Rev. {\bf B54},  R3746 (1996). 
 %
\bibitem{Zotos}
 X. Zotos, W. Lehr, and W. Weber, Z. Phys.  {\bf B74},  289 (1989).
%
%
\bibitem{Haldane}
F.D.M. Haldane,  Phys. Rev. Lett. {\bf 45}, 1358 (1980).
%
\bibitem{Haldane2}
F.D.M. Haldane, J. Phys. {\bf C14},  2585 (1981).
%
\bibitem{Voit}
J. Voit, Rep. Prog. Phys. {\bf 58},  977 (1995).
%
\bibitem{Solyom}
J. Solyom, Adv. Phys. {\bf 28},  209 (1979).
%
\bibitem{Schulz}
H. J. Schulz,
Phys. Rev. Lett. {\bf 64},  2831 (1990). 
%
\bibitem{Matsunami}
T. Matsunami and M. Kimura, Prog. Theor. Phys. {\bf 91},  453 (1994).
%
\bibitem{Matsukawa}
If we regard the origin of the attraction as the super exchange interaction  $J$ acting between the nearest-neighbor Cu-spins,
its order seems to be consistent with the strength of the effective attraction.
Here, the  exchange energy  is given as 
 $J \simeq \frac{2t_{pd}^4}{\Delta^2} [\frac{1}{U_d}+\frac{2}{2\Delta+U_p}]$, which is estimated to be  $J\simeq 0.3$ with $t_{pd}=1$, $\Delta=2$, $U_p=0$, and $U_d=6.5$.
(See H. Matsukawa  and H. Fukuyama, 
J. Phys. Soc. Jpn. {\bf 58},  2845 (1989).) 
%\bibitem{tag} Fake bibitem.
\end{references}
\end{document}